\documentclass[12pt,preprint]{aastex}

\renewcommand{\cite}{\citealp}

\newcommand{\rrl}{{RR~Lyrae}}

\shorttitle{On the remote cluster NGC~2419}
\shortauthors{Ripepi et al.}

\begin{document}

\title{On the remote Galactic Globular Cluster NGC~2419
\altaffilmark{1}}

\author{
Vincenzo Ripepi,\altaffilmark{2}
Gisella Clementini,\altaffilmark{3}
Marcella Di Criscienzo,\altaffilmark{2}
Claudia Greco,\altaffilmark{3} 
Massimo Dall'Ora,\altaffilmark{2}
Luciana Federici,\altaffilmark{3}
Luca Di Fabrizio,\altaffilmark{4}
Ilaria Musella,\altaffilmark{2}
Marcella Marconi, \altaffilmark{2}
Lara Baldacci,\altaffilmark{3}
Marcella Maio,\altaffilmark{3}
}

\altaffiltext{1}{Based on data collected at the 3.5 m Telescopio Nazionale 
Galileo, operated by INAF}

\altaffiltext{2}{INAF, Osservatorio Astronomico di Capodimonte, 
via Moiariello 16, I-80131 Napoli, Italy,
(ripepi, dicrisci, dallora, marcella, ilaria)@oacn.inaf.it}

\altaffiltext{3}{INAF, Osservatorio Astronomico di
Bologna, via Ranzani 1, I-40127 
Bologna, Italy;
(gisella.clementini, claudia.greco, luciana.federici)@oabo.inaf.it}

\altaffiltext{4}{INAF, Centro Galileo Galilei \& Telescopio Nazionale Galileo, PO Box 565, 38700 S.
Cruz de La Palma, Spain; difabrizio@tng.iac.es}

\begin{abstract}

We present a new, deep (V $\sim$ 26) study of the 
Galactic globular cluster NGC~2419 based on 
$B,V,I$ time-series CCD photometry over about 10 years 
and extending 
beyond the cluster published tidal radius.
We have identified 101 variable stars of which 60 are new discoveries, doubling
the known RR Lyrae stars and including 12 SX Phoenicis stars.
The average period of the RR Lyrae stars ($\left<Pab \right>$=0.662 d, 
and 
$\left<Pc\right>$=0.366 d, 
for fundamental-mode $-$RRab$-$ and 
first-overtone pulsators, 
respectively), and 
the position 
in the period-amplitude diagram both 
confirm that 
NGC~2419 is an Oosterhoff
II cluster. The average apparent magnitude of the RR Lyrae stars 
is ${\rm \langle V\rangle}$=20.31$\pm 0.01$ ($\sigma$=0.06, 
67 stars) and leads to the distance modulus 
$\mu_{0}$=19.60 $\pm$ 0.05.
The Color-Magnitude Diagram,
reaching about 2.6 mag below the cluster turn-off, 
does not show clear 
evidence of multiple stellar populations. Cluster stars 
are found until $r\sim 10.5^{\prime}$, and possibly as far as $r\sim 15^{\prime}$,
suggesting that the literature tidal radius might be underestimated. 
No extra-tidal structures are clearly detected in the data.
NGC~2419 has many 
blue stragglers and a well populated
horizontal branch extending 
from the RR Lyrae stars
down to an extremely blue tail ending with the ``blue-hook",
for the first time
recognized in this cluster. The red giant branch is narrow ruling out 
significant metallicity spreads.
Our results seem to disfavor the interpretation 
of NGC~2419 as either having an extragalactic origin or being the relict of a 
dwarf galaxy tidally disrupted by the Milky Way.

\end{abstract}


\section{Introduction}
NGC~2419 is one of the most distant and luminous globular clusters (GCs) in the Milky Way
(MW). Although both the distance ($R_{\rm GC}$ $\simeq$ 90 kpc, \citealt{harris97}) and the 
dynamical parameters (core radius $r_c \sim 9$ pc, and half-mass radius 
$r_h \sim$ 19 pc, \citealt{harris97}) put NGC~2419 among the outer halo Galactic globulars, the 
cluster  
has several unusual properties for an outer halo GC. It is much more luminous and metal-poor 
than the other outer halo clusters: 
with 
$M_V \sim -9.5$ mag (\citealt{harris}) NGC~2419 is among the five brightest clusters in the MW; 
and with 
[Fe/H]$\sim -$2.1 dex (\citealt{suntzeff}) it belongs to the most metal-poor 
group of MW GCs 
that all, except AM-4, are located  
within $R_{\rm GC}$ $\simeq$ 20 kpc.
The cluster horizontal branch (HB)
also resembles that of much closer ``canonical'' metal-poor clusters like M15 or M68, and 
previous investigations 
show that NGC~2419 has the same age of 
M92, within 1
Gyr \citep{harris97}. 
However, NGC~2419
is not an inner halo cluster 
migrated out on an elliptical orbit, since its dynamical 
parameters and orbital properties (\citealt{vdb93,vdb95}, and references therein)
are typical of an outer halo cluster.
NGC~2419 is also anomalous 
in the half-light radius ($R_h$) vs $M_V$ plane (see Fig.~11 of \citealt{mackey05}). 
Among the MW GCs only $\omega$ Cen and M54
have similar properties in this plane, and they 
both are ``peculiar'', since $\omega$ Cen hosts multiple stellar populations 
(see e.g. 
\citealt{bedin04}; \citealt{rey}; \citealt{sollima}) and likely
is the  
stripped core of a defunct dwarf galaxy (\citealt{villanova07}, and references therein); and  
M54 is thought to be the core of the Sagittarius (Sgr) dwarf spheroidal galaxy (dSph) 
that is currently 
merging with the MW
(see e.g. \citealt{layden}). All these peculiarities and the similarity with
 $\omega$ Cen and M54  
suggest that NGC~2419 could have an extragalactic origin and be the relict 
of a dwarf galaxy tidally disrupted by the MW \citep{mackey04}.  
\citet{newberg} 
find that the cluster appears to lie within an overdensity of
 A-type stars connected to
previously discovered 
tidal tails of the Sgr dSph, and conclude that the cluster might once have  
been associated to Sgr. 
In addition, the cluster has a central velocity dispersion ($\sigma_0$)
much lower than the dSphs
for which this quantity has been
measured, 
and,  
in the $\log \sigma_0$ vs $M_V$ plane (\citealt{faber}), 
lies  
3 and 6$\sigma$ apart from the  
``fundamental plane" relations for GCs and elliptical galaxies, respectively 
(see \citealt{degrjis}).
For a comparison, in this plane $\omega$ Cen lies at the intersection 
of these two lines.

Color-Magnitude Diagrams (CMDs) 
of NGC~2419 published so far either do not go fainter than the main sequence turn-off
(TO) or cover small portions of the cluster
(\citealt{christian}, \citealt{harris97}, \citealt{stetson98,stetson05,saha,sirianni}).
We also lack a modern study of the cluster variable stars based on accurate CCD photometry, 
the most recent variability survey  
being the 
photographic work by \citet[][ hereafter PR]{PR} who detected
41 variables 
 in the external regions of NGC~2419 
and found 
the average period of 
the fundamental-mode RR Lyrae stars (RRab) to be 
consistent with NGC~2419 being 
an Oosterhoff II (OoII) cluster (\citealt{oo39}).

In this Letter we present a  
CMD reaching about 2.6 mag below the NGC~2419 TO and a new 
study of the
variable stars based on image subtraction techniques (\citealt{alard2}),
using $B,V,I$ time-series 
CCD photometry covering an area that extends 
well beyond the cluster published tidal radius ($r_t=8.74^{\prime}$,
according to 
\citealt{trager95}). 
The new data are used to verify whether 
multiple stellar populations and tidal tails exist in the cluster 
and to check whether the properties of the RR Lyrae stars
support an extragalactic origin for NGC~2419.

\section{Observations and data analysis}

Time-series $B,V$ photometry of NGC~2419 (RA=07:38:24.0, DEC=38:54:00, J2000)
was collected between
2003 September and 2004 February  with DOLORES at the 3.5m TNG 
telescope\footnote{http://www.tng.iac.es/instruments/lrs/}. 
The TNG data were complemented by WFPC2@HST F555W and F814W archival photometry 
spanning 7 years from 1994
to 2000,
 and by $V,I$ images of the cluster obtained with the Suprime-Cam of the SUBARU 8.2m
telescope\footnote{http://www.subarutelescope.org/}
along four nights in 2002.
The SUBARU dataset covers a total area of 50 $\times$ 43 arcmin$^2$ centered on NGC~2419
and includes both the TNG and HST fields.
Results presented in this Letter refer to a region extending $\pm 10.5^{\prime}$ in North-South and
$\pm 18^{\prime}$ in East-West from the cluster center.
The total number of phase points of the combined datasets reaches  
20, 205 and 48 in the $B$, $V$ and $I$ bands,
respectively, with optimal sampling of the $V$ light curves of RR Lyrae stars,
acceptable coverage in $B$, and  
rather poor sampling in $I$, since the $I$-band images were taken much more
closely spaced in time.  

Images were pre-reduced following standard techniques (bias
subtraction and flat-field correction) with IRAF\footnote{IRAF is
  distributed by the National Optical Astronomical Observatories,
  which are operated by the Association of Universities for Research
  in Astronomy, Inc., under cooperative agreement with the National
  Science Foundation}.
We measured the star magnitudes by PSF photometry running  
 the DAOPHOTII/ALLSTAR/ALLFRAME packages \citep{stetson87,stetson94} on 
 the TNG, HST and SUBARU datasets, separately. 
 Typical internal errors of the $V$ band photometry for
single phase points at the level of the HB are in the range from 0.01 to
0.02 mag.
The absolute photometric calibration was obtained by using local
standards in NGC~2419 from P.B.
Stetson's list\footnote{Available at http://cadcwww.dao.nrc.ca/standards/}. 
Zero point uncertainties 
are of 0.022, 0.014 and 0.014 mag
in $B$, $V$ and $I$, respectively.  
Further details on the data reductions can be found in Di Criscienzo et 
al. (2007, in preparation).

Candidate variable stars were identified using two independent
methods: the Optimal Image Subtraction Technique and the package
ISIS 2.1 
(\citealt{alard2}), applied to the TNG $V$ time-series; and an ad hoc 
procedure applied to the SUBARU $V$ data that included  
calculation of the Fourier transform (in the
\citealt{sc96} formulation) for each star with more than 25 epoch data; evaluation of the
signal-to-noise ratio 
and then analysis of the stars with S/N $>$ 6 and magnitude
$V<$ 23.7 mag (excluding TO and sub giant branch stars).
The two procedures returned a catalogue of 101 confirmed variables.
Periods (and type classification) 
were derived using GRaTiS (Graphical Analyzer of Time Series), 
a custom software developed at the Bologna Observatory (see \citealt{df99,clementini00}).
Precision of the period determinations is of 4-5 decimal places (for variables with periods shorter
than 2 d, 95 objects) and increases up to 6 digits 
for stars with the three datasets (TNG, HST and SUBARU) 
available.
The good sampling of the $V$ light curves allowed a very accurate definition of the star's visual  
mean magnitudes and amplitudes.
Coverage of the 
$B$ and $I$ light curves is generally much poorer, and we often 
estimated average
$B$ and $I$ magnitudes by scaling down in amplitude the star $V$ light curve
to fit observations in the
other bands (see Di Criscienzo et al. 2007). 

\begin{figure}
\includegraphics[scale=.60,angle=-90]{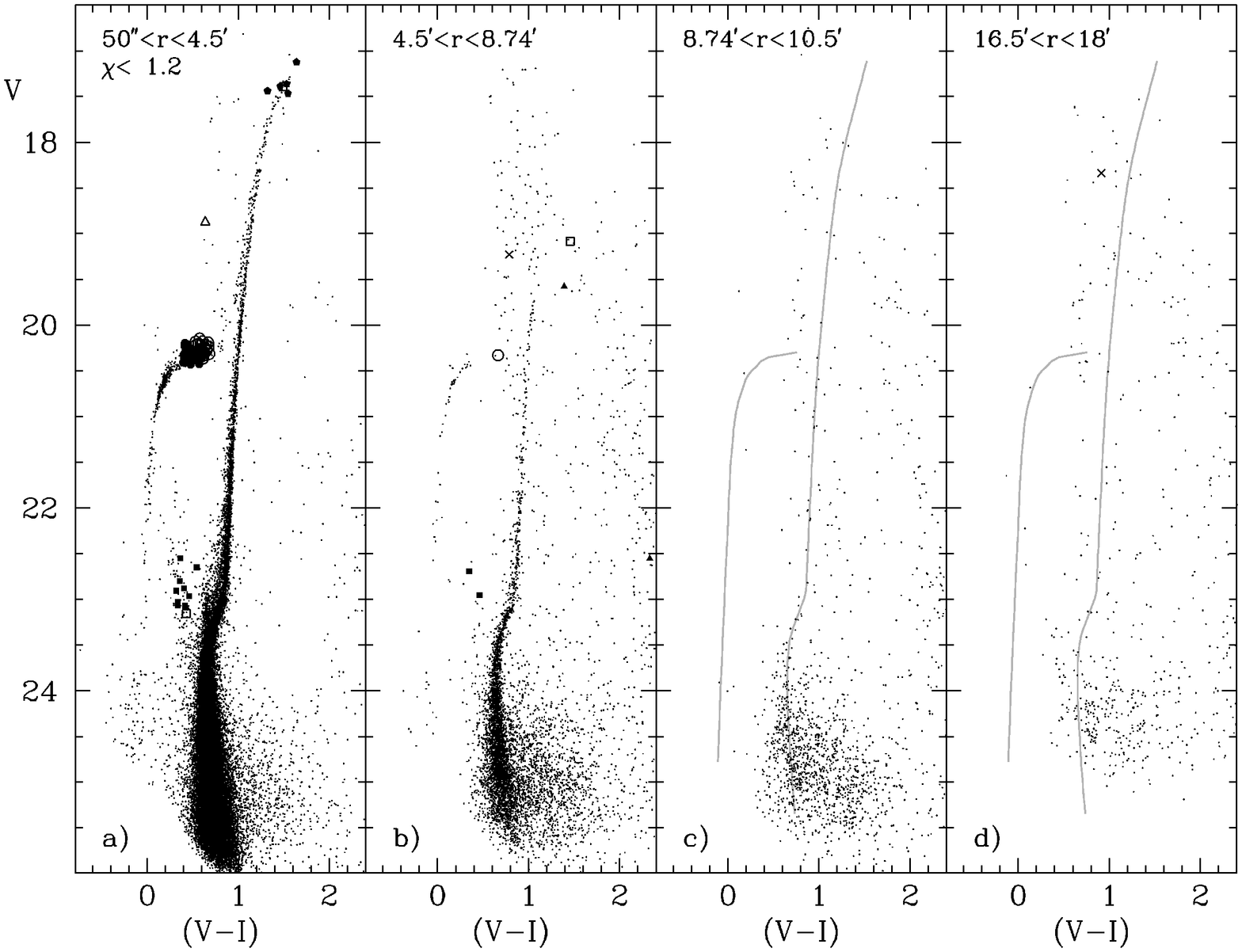}
\caption{$V$, $V-I$ CMD of NGC~2419 in four annular regions at increasing
distance from the cluster center (see labels),  
with the variable stars plotted in different symbols. Open circles:  
{\it ab-}type RR Lyrae stars; filled circles: first overtone and double-mode pulsators; open
triangle: Population II Cepheid; filled squares: SX Phoenicis stars; open squares: binary 
systems; pentagons: long period and semiregular variables; X signs: $\delta$ Scuti stars; 
filled triangles: variables of unknown type. Grey lines in panels {\it c}, {\it d}
are the cluster ridge lines drawn from the CMD in panel {\it a}.}
\label{fig:1}
\end{figure}

Fig.~\ref{fig:1} 
shows the $V$, $V-I$ CMDs (based on the SUBARU dataset) of objects   
 in four annular regions at 
 increasing distance from the cluster center: 50$^{\prime \prime}$ $<$ r $<$ 4.5$^{\prime}$ (panel {\it a}, 
 36262 objects, 91 variable stars); $4.5^{\prime} < r < 8.74^{\prime}$ (panel {\it b}, 
 6116 objects, 
 7 variables); $8.74^{\prime} < r < 10.5^{\prime}$ (panel {\it c}, 1592 objects, zero variables); 
 and 16.5$^{\prime} <$ r $<$ 18$^{\prime}$ (panel {\it d}, 594 objects, 1 variable: a field 
 $\delta$ Scuti star), whose areas are in the ratio 1:3:2:1. Only objects with 
$\chi <$ 1.2, $\sigma_V$ and $\sigma_I \leq$ 0.2 mag, are displayed. 
Panel (d) shows the CMD of an external field devoid of cluster stars with same area as 
the cluster region in panel (a) and thus provides an indication of the contamination 
 by field stars in panel (a).

\section[]{The Variable Star Population}

PR identified 41 variable stars in their photographic study of NGC~2419: 
25 {\it ab-} and 7 {\it c-}type RR
Lyrae stars, 1 Population II Cepheid, 4 red
irregular/semiregular variables, and further 4 variables for which
they did not provide period and classification. PR V39 was later 
recognized as double-mode RR Lyrae star \citep{clement90}. 
We recovered and derived reliable periods for all
the previously known variable stars in NGC~2419, and detected
60 new variables that are mainly located in the cluster central
regions. The new variables include: 11 $ab-$ and 28 $c-$type RR Lyrae stars, 
12 SX Phoenicis stars, 3 binaries, 1 long period
variable (LPV) near the red giant tip, 2 field $\delta$ Scuti stars, and 3 variables
of unknown type. 
Light curves for different types of variables are shown in 
Fig.~2.
The high performance of the image subtraction to detecting small 
amplitude variables in very crowded fields allowed us 
the discovery of a remarkably large number (28 objects) of {\it c-}type RR Lyrae stars
increasing 
fivefold the statistics of 
the RRc stars previously known in NGC~2419. 
Most of the new RRc's as well as the new RRab stars with longer periods were missed by
PR due to their
small
amplitude and the location towards the cluster center. 
Addition of the new discoveries, brings the number of RR Lyrae stars in NGC~2419
to 38 RRab, 36 RRc and 1 RRd star, and  changes 
the ratio of number of RRc over 
number of RRc+RRab stars from 0.28 (PR) to 0.49, in much better agreement 
with typical values of the OoII clusters. 
We also revise the cluster HB morphology parameter HBR (\citealt{lz90}, and reference therein)
from 0.86 (\citealt{harris}), to about 0.76 (after decontamination by field
stars, see Di Criscienzo et al. 2007). 
Average periods are: $\left<Pab \right>$=0.662 d ($\sigma$=0.055,
average on 38 stars), and $\left<Pc\right>$=0.366 d, ($\sigma$=0.038, average on 36 stars), 
for fundamental-mode and first-overtone RR Lyrae stars, respectively, to compare with
$\left<Pab\right>$=0.654 d and $\left<Pc\right>$=0.384 d from PR. The new 
values confirm and strengthen the classification of NGC~2419 as OoII cluster. The 
minimum period of the RRab stars, $Pab_{min}$=0.576 d (star V40, for which PR could not  
derive period information), is 
in perfect agreement 
with values of prototype OoII clusters like M15 and M68. 
Fig.~\ref{f:fig3} shows the $V$-band period-amplitude distribution (Bailey diagram)
of the NGC~2419 RR Lyrae stars. The bulk of RRab variables lies 
in the region occupied by the OoII GGCs (dot-dashed line, from \citealt{clement00}), 
confirming the 
OoII nature of NGC~2419, 
the few stars 
between Oo I and II lines being entirely accounted for by 
crowded objects and possible undetected Blazkho variables (\citealt{blazhko1907}; see discussion in 
Di Criscienzo et al. 2007). For a comparison, we also show in Fig.~\ref{f:fig3} the
period-amplitude distributions of the
{\it bona fide} regular ({\em solid curve}) and well-evolved ({\em dashed curve}) 
{\it ab} \rrl\ stars in M3 
from \citet{cacciari05}.
The average apparent magnitude of the NGC~2419 RR Lyrae stars is 
${\rm \langle V(RR)\rangle}$ = 20.31$\pm 0.01$ ($\sigma$=0.06, average on 67 stars)
excluding objects contaminated by companions.
This value, combined with the cluster
metallicity ([Fe/H]=$-$2.1 dex, \citealt{suntzeff}) and reddening, is used 
to estimate the cluster distance.
\citet {harris}
reports $E(B-V)=0.11\pm0.01$ mag (from the average of various sources), 
while a lower value $E(B-V)=0.065\pm0.010$ is derived on the basis of the \citet{schlegel} maps.
We find $E(B-V)=0.08\pm0.01$ mag by matching the edges of the RR Lyrae instability strip 
of NGC~2419 to those of M68 (\citealt{walker94}, $E(B-V)_{M68}=0.07\pm0.01$) and M5 (\citealt{reid}, 
$E(B-V)_{M5}=0.020\pm0.01$ mag),  
in good agreement with \citet{schlegel}.
Assuming for the absolute luminosity of
the RR Lyrae stars at [Fe/H]=$-1.5$, $M_V$=0.59$\pm$0.03
(\citealt{cc03}), $\Delta M_V/[Fe/H]$=0.214 ($\pm$ 0.047) mag/dex
\citep{clementini03}
for the slope of the
luminosity metallicity relation, $E(B-V)$=0.08 mag, and [Fe/H]=$-2.1$
dex, the distance modulus of NGC~2419 derived from the mean luminosity of
its RR Lyrae stars is: 
$\mu_{0}$=19.60 $\pm$ 0.05 (D= 83.2 $\pm$ 1.9 kpc).

NGC~2419 hosts many blue straggler stars (BSS). Among them we detected  
1 binary system and 12 pulsating variables with periods in the range from 0.041 to 0.140 d 
and several secondary periodicities. They are likely cluster SX Phoenicis stars (see Di Criscienzo 
et al. 2007). One of them is located at $r \simeq 15^{\prime}$ from the cluster center.

\section[]{The Color-Magnitude Diagram}
Fig.~\ref{fig:1} shows that the bulk 
of the NGC~2419 stars and variables is within $r \leq $ 4.5$^{\prime}$ from the 
center (panel {\it a}). 
Cluster
stars 
are found beyond the published tidal radius 
(8.74$^{\prime}$, \citealt{trager95}) until $r \simeq 10.5^{\prime}$ (see panel {\it b}) and 
possibly as far as $r \simeq 15^{\prime}$.
The CMD appears to be dominated by field stars and contaminating galaxies
for $r > 15^{\prime}$. 
The main features of the NGC~2419 CMD are (panels {\it a, b}):
a well defined Main-Sequence reaching about 2.6 mag below the cluster TO 
at $V\sim$ 23.4 mag, with no clear evidence of multiple substructures.
The red giant branch (RGB) is
narrow, ruling out significant differences in
composition among cluster stars. The HB is well
populated to the blue and
has very few 
stars redder than the RR Lyrae.
Most striking
features 
are the well defined BSS sequence outlined by many 
SX Phoenicis stars,
and the extremely prolonged HB blue tail, extending 
down to  $V\sim$ 24.7 mag, (i.e. $M_V\sim5$) with a gap between $V\sim$ 23.4 and 23.8 mag. 
This group of extremely  
hot stars, cannot be fitted by theoretical models 
of extreme HB stars (e.g. \citealt{pietrinferni} post-Helium flash theoretical 
models\footnote{Available at http://www.te.astro.it/BASTI/index.php} 
for the proper composition of NGC 2419) since ``blue-hook" stars 
experience the Helium flash after leaving the RGB.
First detected by  
\citet{whitney98} and \citet{dcruz2000} in their analysis of the
$\omega$Cen HB, ``blue-hook'' stars have been found so far only in very few GGCs 
($\omega$ Cen, M54, NGC 2808, NGC6388, NGC6273, see \citealt{rosenberg}, \citealt{momany}).

\section[]{Discussion and Conclusions}

The results presented in this Letter provide some 
constraints on the hypotheses put forward for the
origin of NGC~2419.
The finding that the cluster confirms to be of 
OoII type,
like ``normal'' low 
metallicity GGCs as M15 and M68, makes its extra-galactic origin unlikely, 
since GCs in external galaxies generally
have properties intermediate between OoI and II types (\citealt{catelan05}).  
The lack of multiple populations or metallicity 
spreads 
does not corroborate 
the hypothesis that NGC~2419 might be the core of a defunct galaxy either. 
In addition, no cluster extra-tidal 
structures are clearly seen in our data beyond $r \sim 15^{\prime}$ to support \citet{newberg}
suggestion of a past association 
with the Sgr dSph, a claim also
disfavored by the Sgr field and cluster (M54) RR Lyrae stars 
having properties on the long-period tail of the OoI group and intermediate
between Oo types, respectively (\citealt{cseresnjes}; \citealt{colucci}).
From our data NGC~2419 appears indeed a very normal, low metallicity Galactic GC,  
the only exception being the HB ``blue-hook'', 
a feature detected so far only in very few globulars and, most noteworthy, 
in those showing multiple main sequences and/or of likely extragalactic origin: 
NGC~2808 (\citealt{piotto07}); $\omega$ Cen (\citealt{villanova07}); and  M54 
(\citealt{layden}).
However, 
the only property these  
clusters seem to share
is a large integrated luminosity (\citealt{rosenberg}). 
On the other hand, 
NGC~2419 peculiar position on the $\log \sigma_0$ vs $M_V$ 
and $R_h$ vs $M_V$ planes remains unexplained. 
{\it ``Clearly there are still all kinds of mysteries that we do not
yet understand at the intercept between dwarf spheroidal
galaxies and globular clusters"} (Sidney van den Bergh 2007, private 
communication).

\acknowledgments 
Financial support for this study was provided by MIUR, under the scientific
projects 2004020323, (P.I.: M. Capaccioli) and PRIN-INAF 2005 (P.I.: M. Tosi).


\clearpage

\clearpage

\begin{figure}
\includegraphics[scale=.80]{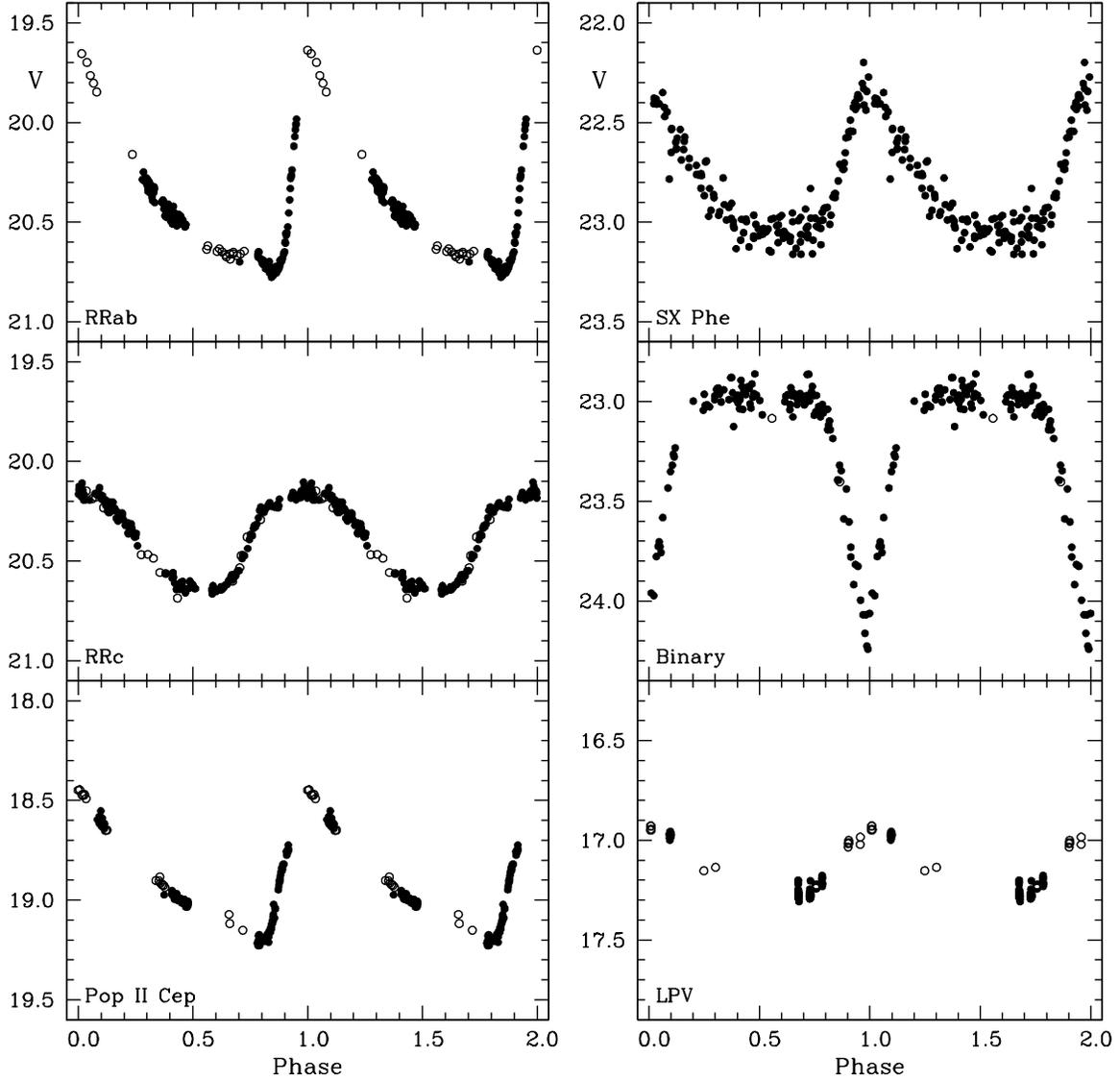}
\caption{$V$ light curves of different types of variable stars in NGC~2419. From top to bottom,
left panel: {\it ab-}, {\it c-}type RR Lyrae stars, Population II Cepheid; right panel:
SX Phoenicis star, binary system, long period variable. Filled and open circles are SUBARU and
TNG data, respectively.}
\label{f:fig2}
\end{figure}
\clearpage

\begin{figure}
\includegraphics[scale=.85]{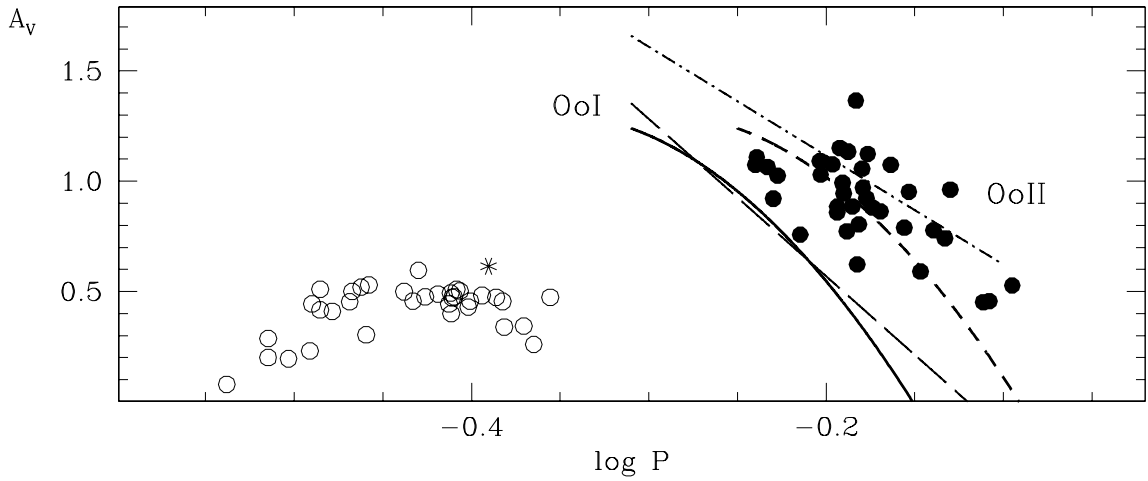}
\caption{ $V$-band period-amplitude diagram for RR Lyrae stars in NGC~2419. Filled and open circles 
are {\emph ab-}
and {\emph c-type} \rrl\ stars, respectively. The asterisk is the double-mode star. 
The {\em straight lines}  
are the positions of the Oosterhoff type I (OoI) and II (OoII) Galactic GC's according to 
\citet{clement00}.
Period-amplitude distributions of the
{\it bona fide} regular ({\em solid curve}) and well-evolved ({\em dashed curve}) 
{\it ab} \rrl\ stars in M3 
from \citet{cacciari05} are also shown for a comparison.
}
\label{f:fig3}
\end{figure}

\end{document}